\documentclass{aa}
\usepackage{natbib}
\bibpunct{(}{)}{;}{a}{}{,}

\usepackage{epsfig}
\usepackage{graphics}
\usepackage{float}
\usepackage{amsmath}
\usepackage{multirow}
\usepackage{longtable}
\usepackage{rotate}
\usepackage{amssymb}
\usepackage{array}
\def\sidehead#1{\noalign{\vskip 1.5ex}\multicolumn{4}{@{}l}{\em #1}\\
                \noalign{\vskip .5ex}}

\def\phn{\phantom{0}}  % Phantom numeral for aligning columns  in tables         
\def\phs{\phantom{$-$}}    % Phantom minus sign for columns in tables
\def\tablecomments#1{\par\smallskip\noindent Notes. #1}
\def\plotone#1{\centerline{\psfig{figure=#1,width=\hsize,clip=}}}
\def\kms{\ifmmode{\rm km\,s^{-1}}\else\hbox{$\rm km\,s^{-1}$}\fi}
\def\nodata{\phs$\cdots$}

\newcommand{\Teff}{$T_{\mathrm{eff}}$}

\let\arcdeg\degr

\let\simlt\la
\setlongtables

\begin{document}

   \title{A new look at the pulsating DB white dwarf GD 358:
       Line-of-sight velocity measurements and constraints
       on model atmospheres 
       \thanks{The data presented herein were obtained
       at the W.M. Keck Observatory, which is operated as a scientific
       partnership among the California Institute of Technology, the
       University of California and the National Aeronautics and Space
       Administration. The Observatory was made possible by the generous
       financial support of the W.M. Keck Foundation.}}

   \author{R. Kotak\inst{1,2}
          \and{M. H. van Kerkwijk}\inst{3,4}
           \and{J. C. Clemens}\inst{5}\thanks{Alfred P. Sloan Research Fellow}
           \and{D. Koester}\inst{6}
           }
   \offprints{R. Kotak}

   \institute{ Lund Observatory
               Box 43, SE-22100 Lund, Sweden \email{rubina@astro.lu.se}
           \and
               Imperial College of Science, Technology, and Medicine,
               Blackett Laboratory, Prince Consort Road, London, SW7 2BZ,
               U.K. \email{rubina@ic.ac.uk}
           \and
              Astronomical Institute, Utrecht University,
              P. O. Box 80000, 3508~TA Utrecht, The Netherlands \\
              \email{M.H.vanKerkwijk@astro.uu.nl}
           \and
              Department of Astronomy and Astrophysics, University of Toronto,
              60 St George Street, Toronto, Ontario M5S 3H8, Canada
              \email{mhvk@astro.utoronto.ca}
           \and
              Department of Physics and Astronomy, University of
              North Carolina, Chapel Hill, NC 27599-3255, USA\\
              \email{clemens@physics.unc.edu}
            \and 
              Institut f\"{u}r Theoretische Physik und Astrophysik, 
              Universit\"{a}t Kiel, 24098 Kiel, Germany\\
              \email{koester@astrophysik.uni-kiel.de} 
            }

   \date{Received 16 September 2002\,/Accepted 30 October 2002}

   \abstract{We report on our findings of the bright, pulsating, helium atmosphere white 
 dwarf GD 358, based on time-resolved optical spectrophotometry. We identify 5 real pulsation 
 modes and at least 6 combination modes at frequencies consistent with those found in previous 
 observations. The measured Doppler shifts from our spectra show variations with amplitudes 
 of up to 5.5\,\kms\ at the frequencies inferred from the flux variations. We conclude that 
 these are variations in the line-of-sight velocities associated with the pulsational motion. 
 We use the observed flux and velocity amplitudes and phases to test theoretical predictions 
 within the convective driving framework, and compare these with similar observations of the 
 hydrogen atmosphere white dwarf pulsators (DAVs).
 The wavelength dependence of the fractional pulsation amplitudes (chromatic amplitudes) allows
 us to conclude that all five real modes share the same spherical degree, most likely, 
 $\ell=1$. This is consistent with previous identifications based solely on photometry.
 We find that a high signal-to-noise mean spectrum on its own is not enough to determine 
 the atmospheric parameters and that there are small but significant discrepancies between 
 the observations and model atmospheres. The source of these remains to be identified.
 While we infer \Teff$\thickspace=24$\,kK and $\log g\sim8.0$ from the mean spectrum, 
 the chromatic amplitudes, which are a measure of the derivative of the flux with respect 
 to the temperature, unambiguously favour a higher effective temperature, 27\,kK, which is 
 more in line with independent determinations from ultra-violet spectra. 
   }
   \authorrunning{Kotak}
   \titlerunning{A new look at GD 358}
   \maketitle
%________________________________________________________________

\section{Introduction}
\label{sec:intro}

Given the potential for successful asteroseismology, pulsating white 
dwarfs have received considerable attention in recent years.
However, this asteroseismological potential has only been tapped 
for a few objects partly due to a lack of understanding of the 
observed behaviour of the pulsations. Turning the problem around, 
the potential to gain insight into the the physics of the upper 
atmosphere of white dwarfs, using the pulsations themselves, is almost 
as great. This approach will ultimately elucidate the nature of the 
pulsations in a variety of ways.

White dwarfs can be broadly divided into two groups: those with
atmospheres composed almost exclusively of hydrogen (DA types),
and those with nearly pure helium atmosphere (DB types). 
The former comprise up to 80\% of the total population of white
dwarfs, while the latter dominate the remaining 20\%.

Both groups of white dwarfs evolve through a phase of pulsational
instability; the DB instabilty strip stretches from $\sim$\,27-22\,kK, 
while that of the DAs ranges from $\sim$\,12.5-11.5\,kK. Both are 
non-radial, g-mode pulsators \citep{chan:72,wr:72}, and typical 
pulsation periods are of the order of several hundred seconds. This 
evolutionary phase provides a window into the interiors of these 
objects.

How do the global properties of the pulsating DBs ($\equiv$\,DBVs)
compare with those of the better-studied DAVs? Do they follow the same
trends across the instability strip? Are the same driving and amplitude
saturation mechanisms at work? 

The driving of the pulsations was originally thought to occur in 
an ionisation zone via the $\kappa$-mechanism i.e. in a manner akin
to that of the Cepheids and $\delta$-Scuti type stars 
\citep[e.g.][]{dzk:81,dv:81,wing:82a}. 
However, \citet{brick:83,brick:91} realised that the convective turnover 
time is short ($\sim$\,1\,s) compared with the mode periods, meaning 
that the convective zone can respond instantaneously to the pulsations.
Based on the original ideas of \citet{brick:90,brick:91,brick:92},
\citet{gw:99a,gw:99b}, and \citet{wg:99} have extended the theory of 
mode driving via convection. They detail the trends of measurable 
quantities as a function both of mode period and effective temperature 
i.e. across the instability strip. These trends, originally formulated 
for the DAVs, are expected to be equally applicable to the DBVs. 
While certain aspects of the convective-driving mechanism have been 
confronted with observations for the DAVs \citep[e.g.][]{vkcw:00,kotakhs:02}, 
this has not been the case thus far for the DBVs. We intend to test at 
least some of these predictions.

Although these were previously thought to be too small to measure,
\citep[given the instrumentation available at that time,][]{rkn:82}, 
the line-of-sight velocity variations due to the pulsations have 
recently been measured using time-resolved spectroscopy
by \citet{vkcw:00} for one of the best-studied DA pulsators, \object{ZZ Psc} (a.k.a.
G 29-38). This opened up the possibility of probing the upper atmosphere of 
white dwarf pulsators using an entirely new tool. Since then, velocity variations, 
or stringent upper limits to these, have been measured for about half 
a dozen DAVs \citep[e.g. \object{HS 0507+0434B}, \object{HL Tau 76}, \object{G 226-29}, 
\object{G 185-32},]
[respectively]{kotakhs:02,kotakred:02,thompson:02}. 
Our primary goal is to apply the same technique to a DBV
in order to compare the velocity and flux amplitudes and phases for each mode 
with theoretical expectations, and with previous measurements for the DAVs. 

Our data also hold the possibility of testing model atmospheres by means
of our extremely high signal-to-noise mean spectrum, and by comparing the
wavelength dependence of pulsation amplitudes with those computed using
model atmospheres. A by-product is an independent check on mode identifications,
previously determined from photometry only. We discuss these in reverse order
below. 

Asteroseismology relies on the correct identification of modes present 
in the pulsational spectrum i.e. assigning the radial order ($n$), the 
spherical degree ($\ell$), and the azimuthal order ($m$) to each mode.
This is usually not a trivial process. 
Traditionally, two methods are used to determine the $\ell$ and $n$ 
values of the observed modes. The first is a comparison of the observed 
distribution of mode periods with predicted ones. The second relies on 
observing all (i.e. 2$\ell$+1) rotationally-split multiplets within a 
period group.

The success of both methods has been limited for a number of reasons.
The paucity of observed modes has hindered the use of the first 
method, while the time-base of the observations is often not long enough 
to resolve rotationally-split multiplets; also, different components 
in a multiplet are not always detected even in long time series.
The main stumbling block, however, is a general lack of understanding of 
the cause(s) of amplitude variability in the observed pulsation spectra on 
a variety of seemingly irregular timescales. Clearly, other complementary 
methods for identifying modes are highly desirable.

Within the context of mode identification, \citet{robetc:95} emphasised 
the useful properties of fractional, wavelength-dependent pulsation
amplitudes (``chromatic amplitudes''). This method relies on the 
increased importance of limb darkening at short (ultra-violet) wavelengths 
which has the effect of increasing mode amplitudes in a manner that is a function
of $\ell$. This holds insofar as the pulsations can be described by spherical 
harmonics, and that the brightness variations are principally due to 
variations in temperature \citep{rkn:82}. \citet{cvkw:00} showed that a 
similar effect is also at play at optical wavelengths only.

Quite apart from their use as potential $\ell$-identifiers, chromatic
amplitudes provide an additional constraint to model atmospheres.
Traditionally, the aim has been to reproduce the observed (integrated) flux 
($F$) over a wide wavelength range. Given the insensitivity to temperature 
of Balmer lines and \ion{He}{i} lines at optical wavelengths for the DAVs 
and DBVs respectively, model atmospheres have to be constrained by other 
means. The temperature variations due to the pulsations provide just such 
a constraint, in the form of $dI(\lambda,T,\mu)/dT$ where $T$ is the temperature
and $I$ the intensity, as a function of the limb angle ($\mu$).
For at least one DAV, \citet{cvkw:00} have shown that in spite of obtaining
an excellent fit to the observed mean spectrum, the fits to the observed 
chromatic amplitudes were unsatisfactory.
This could point to inconsistencies in the model atmospheres that result
in, for example, a misrepresentation of the temperature stratification.
Given the intractability of treating convection in models, this may be the 
case even though the model spectra match the observations rather well, 
especially for the DAVs.
Our secondary aim therefore, is to check if such inconsistencies 
are present in DBV model atmospheres and to possibly identify the source 
of these.

This paper is organised as follows:
we begin with a recapitulation of previous Whole Earth Telescope observations and 
effective temperature determinations of GD 358, followed by a description of 
our data reduction.  In Sect. \ref{sec:modatm} we analyse our high signal-to-noise 
average spectrum. We extract various quantities from our light and velocity 
curves in Sect. \ref{sec:ligovel} and compare our measurements with 
expected trends. We examine the chromatic amplitudes in Sect. 
\ref{sec:champ} and conclude in Sect. \ref{sec:conc}. 

\section{GD 358 (V777 Her; WD 1645+325)}
\label{sec:gd358}

\subsection{Pulsations}

Being the brightest DBV ($V=13.65$), GD 358 has been the object of considerable 
scrutiny since the discovery of its variablity about two decades ago \citep{wing:82b}. 
A Whole Earth Telescope \citep[WET;][]{nath:90} run in 1990 revealed a rich pulsation 
spectrum with many multiplets simultaneously excited, allowing the pulsation
models to be well-constrained. 
As all modes displayed clear triplets, and as the period spacing
(i.e. the difference in period between consecutive triplet groups) was
approximately that expected from asymptotic (large radial order ($n$) limit)
theory, all modes were assigned $\ell=1$.
This afforded an asteroseismological determination of fundamental stellar quantities 
and the interior structure; for instance, comparing the observed periods and 
period-spacings to pulsation models implied a mass of $0.61\pm 0.03\,M_{\odot}$, 
while differences in the splittings of modes of the same spherical degree 
($\ell=1$) and successive overtones were best interpreted as being the signature 
of differential rotation \citep{wing:94,bw:94}.

A subsequent WET run in 1994 \citep{vuille:00} revealed a pulsation spectrum that
was qualitatively similar to the one obtained in 1990, but different in the
details: mode variability, both in amplitude and more surprisingly, in frequency, 
was apparent. Furthermore, no modes could be identified as having $\ell>1$, although 
some excess power was clearly present. A large number of harmonics and combination 
modes (linear sums and differences of the real modes) were present in both runs.

\subsection{Atmospheric parameters}

\begin{figure*}[t]
\includegraphics[width=\textwidth]{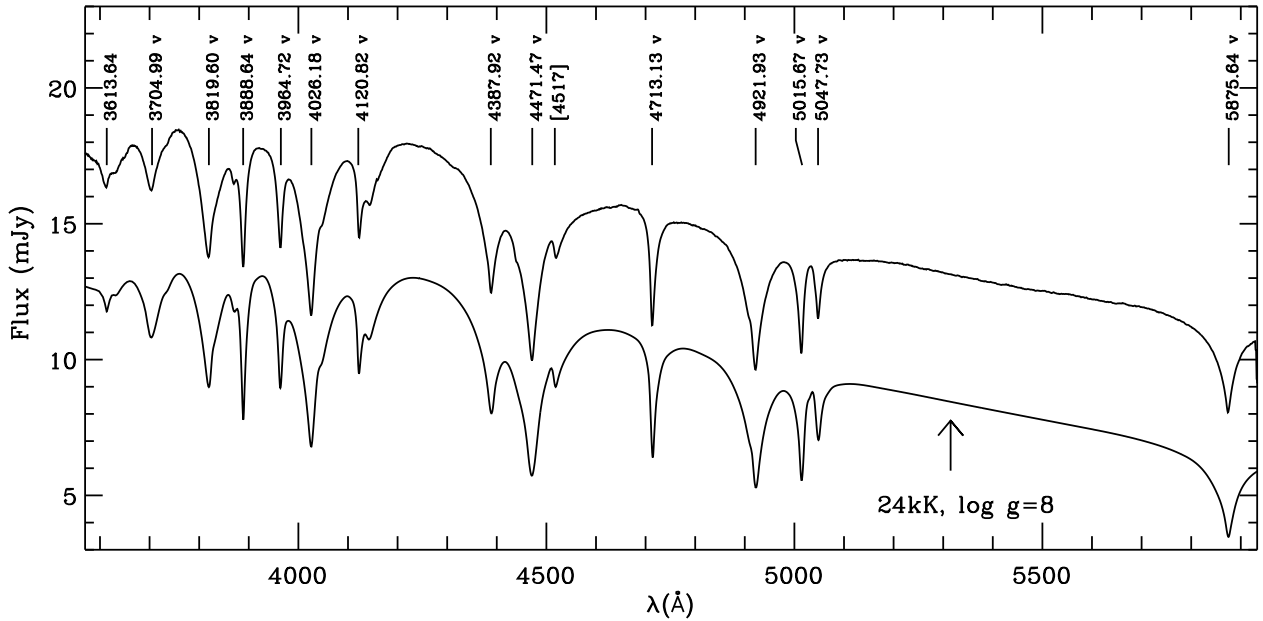}
\caption{Time-averaged spectrum of GD 358 displaying lines of \ion{He}{i} only.
Note the presence of the forbidden line at 4517\,{\AA}, well-separated from its 
allowed component at 4471\,{\AA}. Only the stronger lines are annotated. 
The lower spectrum is a model offset by $-$4.5\,mJy with the parameters 
indicated, and computed using ML2/$\alpha=0.6$.
A ``v'' next to the line label indicates that the Doppler shift of the line 
was used in the construction of the velocity curve.}
\label{fig:avspec}
\end{figure*}

In spite of having been extensively observed, the atmospheric parameters
of GD 358 and of DBs in general, are still debatable and have given rise to 
a not inconsiderable body of literature. This is perhaps not surprising given 
that the location and extent of both the DBV instability strip and the DB 
gap -- the absence of pure helium-atmosphere white dwarfs in the range 
30 -- 45\,kK range -- as well as constraints on the efficiency of convection 
all hinge on accurate effective temperatures of DB white dwarfs as a whole.

The difficulty in accurately determining the effective temperatures of
DBVs stems partly from the fact that at optical wavelengths, the absorption
lines of \ion{He}{i} are relatively insensitive to changes in effective
temperature and surface gravity. While the advent of the {\em International
Ultraviolet Explorer\/} (IUE) satellite meant that the temperatures could
now be estimated near the peak of the energy distributions, it also brought
some puzzling discoveries. The analysis of \citet{liebert:86} based on
IUE data and a variety of model atmospheres set the blue edge of the
instability strip at 32\,kK and the effective temperature of GD 358 at
$27^{+1}_{-2}$\,kK. A reanalysis of the same IUE data by \citet{thejll:90}
using line-blanketed models placed GD 358 at $24\pm1$\,kK. \citet{beau:99} 
investigated the effect of trace amounts of hydrogen in the atmosphere of DBs. 
They found that this increase in continuum opacity could lower the derived 
effective temperatures by up to 3.5\,kK. Their fits to optical spectra of
GD 358 using models with and without hydrogen yielded 24.7\,kK and 24.9\,kK 
respectively which were in good agreement with the proposed mean value of 
$24\pm1$\,kK by \citet{dk:85}.

\begin{table}[!t]
\setlength{\tabcolsep}{5pt}
\caption[]{Summary of previous \Teff\ determinations of GD 358}
\fontsize{7.7}{10}\selectfont
 \begin{centering}
 \begin{tabular}{llll}
 \label{tab:gdteff}
                          &   &                &                        \\
\hline
                          &   & \Teff          &  Comment                \\
                          &   & (kK)           &                          \\
\hline
 \citet{dk:85}            &   & 24$\pm$1        &  optical/IUE spectra    \\
 \citet{liebert:86}       &   & 27$^{+1}_{-2}$  &  IUE                    \\
 \citet{thejll:90}        &   & 24$\pm$1        &  line-blanketed models  \\
 \citet{beau:99}          &  \multirow{2}{-2.3cm}{\{}  & 24.7    &  with H  \\
                          &   & 24.9            &  without H                \\
 \citet{prov:96,prov:00}  &   & 27$\pm$1        &  GHRS                    \\
 \citet{bw:94}            &   & 24$\pm$1        &  Observed period spacing  \\
                          &   &                 &  c.f. $M=0.6\,M_{\odot}$ model \\
\citet{mnw:00}            &   & 22.6   &  best fit to WET periods  \\
                          &   &                 &  from genetic-algorithm-  \\
                          &   &                 &  based optimisation \\
\hline
\end{tabular}
\tablecomments{See text (Sect. \ref{sec:gd358}) for details. This list is not an
exhaustive one.}
\end{centering}
   \end{table}

Meanwhile, \citet{sion:88} reported the intriguing discovery of \ion{C}{ii} 
$\lambda\,1334.75$\,{\AA} and \ion{He}{ii} $\lambda$\,1640.43\,{\AA} in 
IUE spectra of GD 358. The presence of ionised species implies a high 
temperature while the mere presence of carbon has implications for the 
thickness of superficial helium layer and dredge-up of material by the 
convection zone from the interior.

A more recent investigation in the ultra-violet using the Goddard High Resolution 
Spectrograph (GHRS) on board the {\em Hubble Space Telescope\/} by
\citet{prov:96,prov:00} confirmed the presence of carbon in the photosphere 
of GD 358.
Based on the presence of the \ion{He}{ii} $\lambda$\,1640\,{\AA} line, 
\citet{prov:96} concluded that $T_{\mathrm{eff}}=27\pm1$\,kK. Note though,
that as the theoretical calculations were wanting, their line profile was
computed for an {\em ionised\/} plasma.\footnote{\citet{prov:96} used the 
calculations of \citet{schobut:89} who do not consider the effects of
\ion{He}{i} on \ion{He}{ii}.} The absence of the \ion{C}{i} 
$\lambda$\,1329\,{\AA} line together with the presence of the 
\ion{C}{ii} $\lambda$\,1335\,{\AA} line in the observed spectra allows a 
lower limit to be placed on the effective temperature. Based on the same data,
\citet{prov:00} focused on using models to determine the temperature at which 
the \ion{C}{i} line just disappears. Although they could not match the profile 
of \ion{C}{ii} for any combination of temperature and abundance making their 
use of $\chi^{2}$ analysis questionable, they arrive at a value of $27\pm1$\,kK. 
These determinations are only marginally consistent with the systematically
lower ones based on optical spectra.

Interestingly, \citet{bw:94} derive $24\pm1$\,kK from the average period spacing 
of the real modes observed during WET 1990 run together with models appropriate for 
a $\sim\,0.6M_{\odot}$ white dwarf, while \citet{mnw:00} obtain 22.6\,kK
as their best-fit temperature from a genetic-algorithm-based optimisation
technique applied to the same WET periods.
It is evident from Table \ref{tab:gdteff} that uncertainties in the effective
temperature of GD 358 persist. In what follows, we consider temperatures
between 24 and 28\,kK.

Although the determination of the precise effective temperature is dependent on
the convective efficiency used in the model atmospheres, a relative temperature
measurement is more straightforward. Most studies firmly plant GD 358 close 
to the blue edge of the DB instability strip. We describe below the constraints
our data allow us to place on the atmospheric parameters of GD 358.

\section{Data acquisition and reduction}
\label{sec:datared}

\begin{figure*}[!t]
\plotone{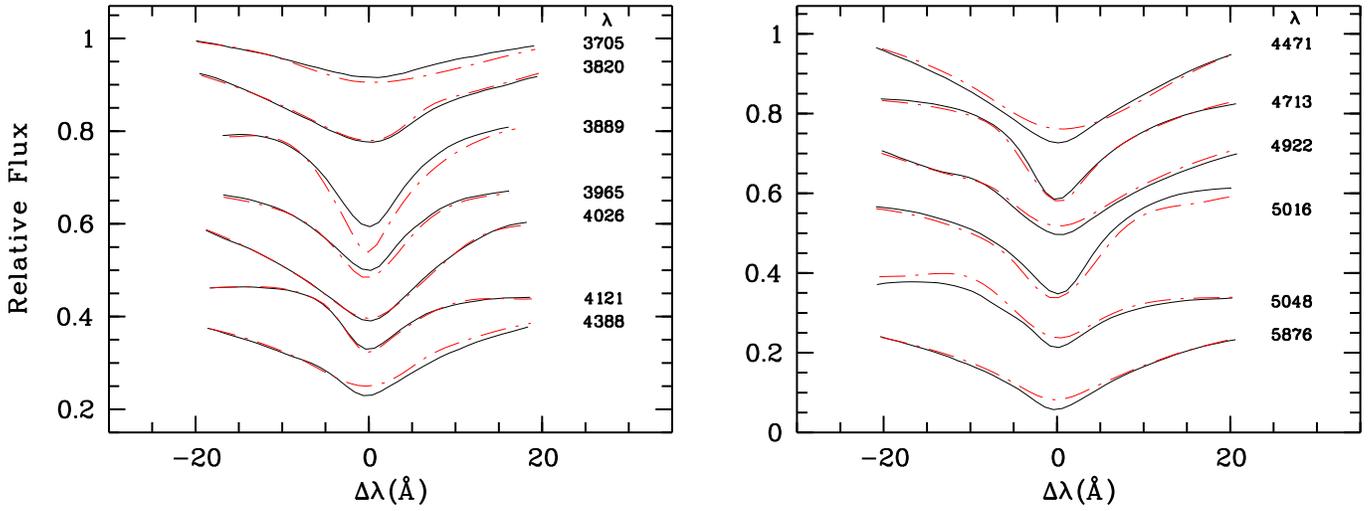,angle=-90}
\caption{Strong lines in the average spectrum of GD 358 (full line) overlaid
 with a model (dot-dashed line) having \Teff = 24\,kK, $\log g=8$, and 
 ML2/$\alpha=0.6$. Both the model and observed line profiles have been
 normalised in the same manner, with each line treated separately as
 described in Sect. \ref{sec:modatm}.} 
\label{fig:stackmodavspec}
\end{figure*}

Time-resolved spectra of GD 358 were obtained on 1999 June 23 using
the Low Resolution Imaging Spectrometer \citep[LRIS,][]{oke:95} mounted
on the Keck II telescope. An 8$\farcs$7-wide slit was used together 
with a 600 line\,mm$^{-1}$ grating covering approximately 3500-6000\,{\AA} 
at 1.24\,{\AA}\,pixel$^{-1}$. The wavelength resolution was approximately 
5\,{\AA}. A set of 612 20\,s exposures were acquired from 06:53:12 to 12:37:55\,U.T., 
bracketed by a series of arc and flatfield frames. The reduction of the data was 
carried out in the MIDAS\footnote{The Munich Image Data Analysis System, developed 
and maintained by the European Southern Observatory.} environment and included the 
usual steps. Additionally, corrections for instrumental flexure and differential 
atmospheric refraction were applied. The procedure is nearly identical to that 
detailed in \citet{vkcw:00}. 

\begin{figure}[htb]
\plotone{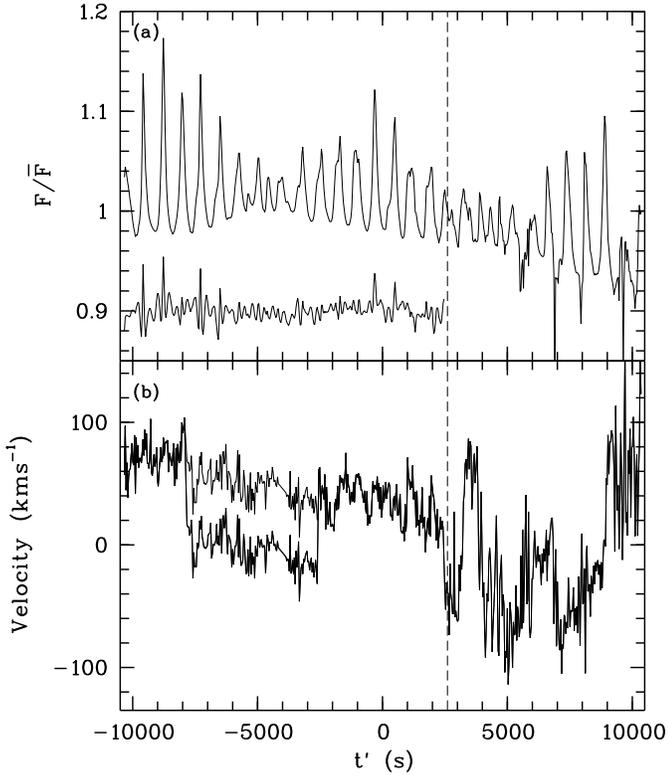}
\caption{\textbf{(a)} Lightcurve of GD 358 showing a typical beat envelope.
Note that the maxima are stronger and sharper than the minima. The lower curve
shows the residuals (offset by +0.90) after fitting sinusoids with the amplitudes
and phases listed in Table \ref{gdtab}. \textbf{(b)} The velocity curve as constucted
by crosscorrelation. The vertical dashed line indicates the point ($t' \gtrsim\ 2463$\,s)
beyond which the remainder of the frames were discarded. This was due to rapidly
deteriorating seeing as discussed in Sect. \ref{sec:datared}. 8 additional frames from
($ -4166 \lesssim t'(\mathrm{s}) \lesssim -3899$) had to be deselected as these were
markedly deviant from the expected mean position. 
The lower portion of the velocity curve between $-7410 \lesssim t'(\mathrm{s}) \lesssim
-2701$ shows the constant offset referred to in Sect. \ref{sec:datared}.}
\label{fig:ligovel}
\end{figure}

The seeing was $0\farcs$9 at the beginning of the run but deteriorated to 
1$\farcs$5 towards the end of the run. This led to guiding problems 
which resulted in significant jitter (up to 2 pixels) of the target in the
slit later in the run. We tried to correct for the jitter in the dispersion
direction using the jitter in the spatial direction (which we could measure
from the spatial profile), but found that the two were unfortunately 
uncorrelated. Not wanting to compromise the accuracy of our Doppler-shift 
measurements, we simply discarded the last 233 frames in the analysis that we 
describe below. Between 07:33:50 and 09:01:49\,U.T. we found a constant offset 
in the pixel positions of the lines. We corrected for this shift using the
mean offset as determined from frames taken before and after the above times.
(see Fig. \ref{fig:ligovel}).

For flux calibration, 11 frames of the flux standard Wolf 1346 were taken,
for which fluxes in 20\,{\AA}-wide bins were available. However,
from previous observations, we have found that the response of the spectrograph
shows small but significant variations over wavelength ranges smaller than
20\,{\AA}, and that these can be removed efficiently by using observations of
stars for which well-calibrated model spectra are available. While these
variations do not affect our analysis of the chromatic amplitudes (as these
cancel out to first order), they do matter for the comparison of the mean 
spectrum with models. We therefore performed the flux calibration in two steps: 
we first calibrated all spectra with respect to the flux standard
G 191-B2B (spectra for which were taken during a separate run but with an
identical set-up) using the calibration of \citet{bohlin:95},
and then derived a (linear) response correction factor by comparing the 
observed spectrum of Wolf 1346 calibrated with respect to G 191-B2B with its 
own calibrated fluxes. This small correction factor was then applied to the 
time-averaged spectrum of GD 358 to ensure that the relative calibration would 
not affect the comparison with model spectra. We further scaled the mean spectrum 
(by a factor of 0.73) such that the estimated flux corresponded to the observed 
V-band magnitude.

\section{Model atmospheres and the mean spectrum}
\label{sec:modatm}

Averaging together all spectra results in a high signal-to-noise mean spectrum 
that shows lines of \ion{He}{i} only (Fig. \ref{fig:avspec}). 

By comparing with a grid of tabulated DB model atmospheres \citep[details in]
[]{fin:97} spanning a range from 18-30\,kK in steps of 500\,K, and $\log g$ values 
from 7.5-8.5 in increments of 0.25 dex, we attempted to derive $T_{\mathrm{eff}}$ 
and $\log g$ from our mean spectrum. The model atmospheres consist of pure helium
and were computed under the assumption of LTE; the description of line-broadening 
was taken from \citet{beau:97}. Convective transport was taken to be of intermediate 
efficiency (ML2/$\alpha=0.6$). The parameter $\alpha$ denotes the mixing length as a 
fraction of the pressure scale height, with larger values of $\alpha$ corresponding 
to more efficient energy transport by convection. Several versions of the mixing length 
(ML) approximation abound. The original concept is due to \citet{bv:58}.  The 
ML2 version referred to above, with $\alpha=1$ is due to \citet{bc:71} and differs 
from the ML1 and ML3 prescriptions in the choice of certain parameters that determine 
the horizontal energy loss rate and thereby influence the convective efficiency.

Our best fit model with \Teff$=24$\,kK and $\log g=8.0$ is shown in Fig. 
\ref{fig:stackmodavspec}. Each line was normalised separately, but in an
identical manner for both the model and observed spectrum. The normalisation
was carried out by dividing by the fit to the continuum points on either side 
of the line, thereby preserving the slope. We estimate the error in \Teff\ to 
be about 0.5\,kK.

Small discrepancies between the model and the observations remain. These would 
normally be masked by a lower signal-to-noise spectrum. However, the results of 
Sect. \ref{sec:champ} imply that these small discrepanices are probably real
rather than being due to inadequacies in the flux calibration.

\section{Light and velocity curves}
\label{sec:ligovel}
\subsection{Light curve}

\begin{figure*}[!t]
\sidecaption
\mbox{
\epsfxsize=0.795\textwidth \epsfbox{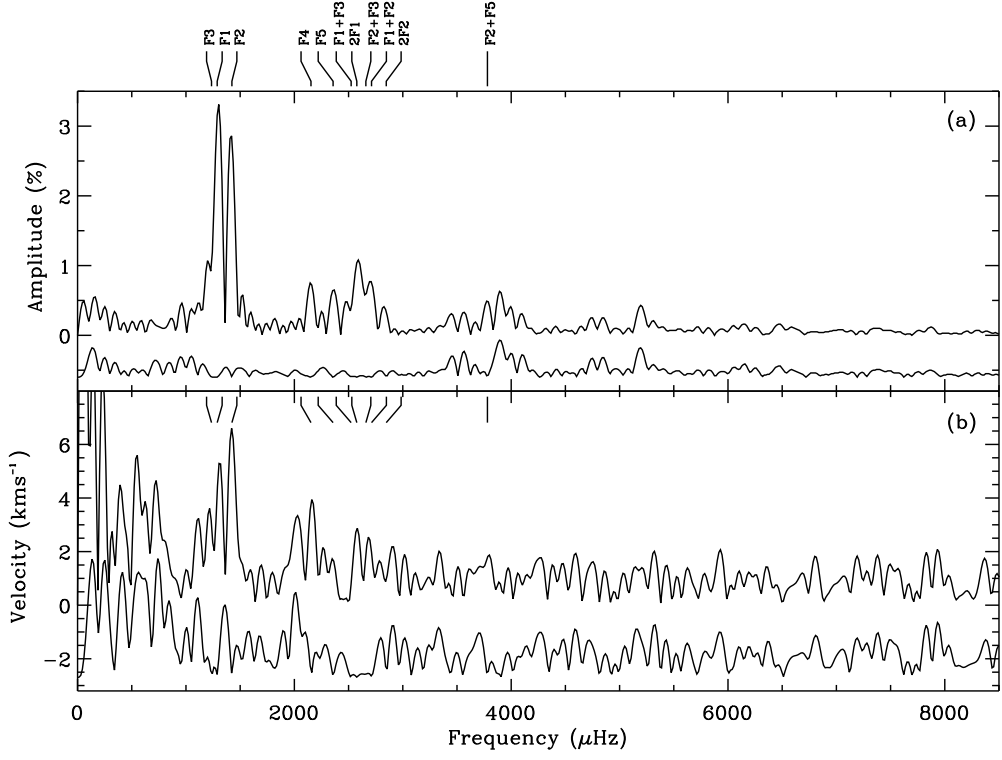}
     }
\caption{\textbf{(a)} Fourier Transform of the light curve and residuals
 offset by $-0.6$\% and \textbf{(b)} of the velocity curve shown up to
 8500\,$\mu$Hz with the residuals offset by $-2.7$\,\kms\. The strongest 
 peaks in the FT of the light curve are labelled. There are no peaks greater 
 than $\sim$\,0.08\% and $\sim$2.7\,\kms\ longward of 8500\,$\mu$Hz.}
\label{fig:flxovel}
\end{figure*}

The lightcurve shown in Fig. \ref{fig:ligovel}a was constructed by dividing the 
line-free region of the spectra between $5100-5800$\,{\AA} by its average. The time 
axis was computed relative to the middle of the time-series ($t'=t-9:35:33$\,U.T.) 
in order to minimise the covariance between the amplitudes ($A$) and phases ($\phi$) 
of the sinusoids we use to fit the light curve. The Fourier Transform, calculated up to 
the Nyquist frequency, is shown in Fig. \ref{fig:flxovel}a. Periodicities were 
determined consecutively in order of decreasing amplitude by successively fitting 
functions of the form $A\cos(2\pi ft'-\phi) + C$ where $f$ is the frequency, and $C$ 
a constant offset. The values we obtain for $f, A,$ and $\phi$ are listed in Table 
\ref{gdtab}. We first identified the real modes (i.e. those that were not obvious
linear combinations) and then used these to fit the combination modes by fixing
the frequency to the sum or difference of the real modes. We additionally imposed 
the requirement that the combination mode have an amplitude smaller than those of 
the two constituent real modes. 

As is obvious from the residuals (Figs. \ref{fig:ligovel}a, \ref{fig:flxovel}a), the 
light curve is not free of periodicities after subtracting the frequencies listed in 
Table \ref{gdtab}. We have not attempted to look for lower amplitude modulations not only 
because our short time coverage and consequentially low frequency resolution means that 
the distinction between real and combination modes and noise peaks is blurred, but also 
because the determination of accurate periods and identifying all possible combination 
modes is not the principal aim of this investigation. 

The frequencies that we measure for our real modes agree well with those measured
for the two previous WET runs \citep{wing:94,vuille:00} although detailed comparison
is difficult, as the signal we measure is a blend of the various $m$ components.  
For instance, the peaks at 3886\,$\mu$Hz ($\sim$3\,F1) and 5182\,$\mu$Hz ($\sim$4\,F1) 
are probably due to combinations of unresolved multiplet components.
We list our real modes together with the amplitudes, $n$, and $m$
determinations from the above studies in Table \ref{tab:gdwet}. The $m$-values listed
are the closest match between our periods and frequencies compared to those listed in
Table 3 of \citet{vuille:00} only.

\begin{table*}[!ht]
\caption[]{Pulsation frequencies and other derived quantities from the light
and velocity curves of GD 358.}
\fontsize{8.1}{10}\selectfont
 \begin{centering}
 \begin{tabular}{lcccccccc}
 \label{gdtab}
    &           &                 &           &                &            &                &                  &          \\
\hline
Mode &   Period &  Frequency      & $A_{L}$   &   $\Phi_{L}$   &  $A_{V}$    &    $\Phi_{V}$ & $R_{V}$          & $\Delta\Phi_{V}$ \\
     &    (s)   & ($\mu$\,Hz)     & (\%)      &  (\arcdeg)     &   (\kms)    &   (\arcdeg)   & (Mm\,rad$^{-1})$ & $(\arcdeg)$ \\
\hline
\sidehead{Real Modes:}
 F1 (17) &  776.42 $\pm$ 0.54 &  1288.0 $\pm$ 2.6 & 3.10 $\pm$ 0.18 & $-$128     $\pm$ \phn3  & 4.6 $\pm$ 1.1 & $-$126     $\pm$ 14 & 18 $\pm$ \phn4  & \phs\phn2  $\pm$ 14 \\
 F2 (15) &  702.39 $\pm$ 0.75 &  1423.7 $\pm$ 1.5 & 2.29 $\pm$ 0.08 & $-$153     $\pm$ \phn3  & 5.6 $\pm$ 1.0 & \phn$-$95  $\pm$ 10 & 27 $\pm$ \phn5  & \phs59     $\pm$ 11 \\ 
 F3 (18)  &  809.28 $\pm$ 0.29 &  1235.7 $\pm$ 6.5 & 1.07 $\pm$ 0.17 &	\phs\phn71 $\pm$ \phn9  & 4.0 $\pm$ 1.1 & \phs\phn77 $\pm$ 16 & 48 $\pm$ 14     & \phs\phn5  $\pm$ 18 \\ 
 F4 (9) &  464.44 $\pm$ 0.10 &  2153.1 $\pm$ 5.1 & 0.66 $\pm$ 0.08 & \phs\phn89 $\pm$     10 & 3.3 $\pm$ 1.0 & \phs150    $\pm$ 17 & 37 $\pm$ 11     & \phs61     $\pm$ 20 \\ 
 F5 (8) &  424.26 $\pm$ 0.78 &  2357.1 $\pm$ 4.3 & 0.62 $\pm$ 0.08 & \phs116    $\pm$     10 & 1.9 $\pm$ 1.0 & $-$164     $\pm$ 29 & 21 $\pm$ 11     & \phs81     $\pm$ 31 \\
\sidehead{Combinations:}				    
Mode    &   Period &  Frequency      & $A_{L}$   &   $\Phi_ {L}$ &  $A_{V}$       & $\Phi_{V}$     & $R_{C}$  & $\Delta\Phi_{C}$ \\
        &    (s)   & ($\mu$\,Hz)     & (\%)    &  (\arcdeg)      &  (\kms) & (\arcdeg)             &          & $(\arcdeg)$     \\
2F1    &  388.21 & 2575.9 & 1.12 $\pm$ 0.11 & \phs105    $\pm$ \phn8 & 2.5 $\pm$ 1.1 & \nodata & 12     $\pm$ 2 &  \phs\phn1  $\pm$ 10 \\
F1+F2  &  368.78 & 2711.7 & 0.55 $\pm$ 0.11 & \phs\phn83 $\pm$    11 & 0.8 $\pm$ 1.2 & \nodata & \phn4  $\pm$ 1 &  \phs\phn5  $\pm$ 11 \\
F2+F5  &  264.50 & 3780.8 & 0.52 $\pm$ 0.08 & \phn$-$91  $\pm$    10 & 1.8 $\pm$ 1.0 & \nodata & 18     $\pm$ 4 & $-$53       $\pm$ 14 \\
F2+F3  &  376.03 & 2659.4 & 0.48 $\pm$ 0.09 & $-$101     $\pm$    14 & 1.0 $\pm$ 1.1 & \nodata & 10     $\pm$ 2 &  $-$19      $\pm$ 17 \\
F1+F3  &  396.26 & 2523.6 & 0.48 $\pm$ 0.13 &  \phn$-$40 $\pm$    15 & 0.8 $\pm$ 1.1 & \nodata & \phn7  $\pm$ 2 &  \phs17     $\pm$ 18 \\
2F2    &  351.20 & 2847.4 & 0.31 $\pm$ 0.08 & \phs\phn32 $\pm$    15 & 1.2 $\pm$ 1.0 & \nodata & \phn6  $\pm$ 2 & $-$22       $\pm$ 16 \\
\hline
\end{tabular}
\tablecomments{The number in brackets next to the mode name indicates the $n$
value from previous WET studies, based on their $\ell=1$ identification for all
modes. The velocity to flux amplitude ratio $R_{V} = A_{V}/(2\pi fA_{L})$, and phase
$\Delta\Phi_{V} = \Phi_{V} - \Phi_{L}.$
The photometric amplitude for combination modes with respect to the real modes that make
up the combination is given by $R_{C} = A_{L}^{i\pm j}/(n_{ij}A_{L}^{i}A_{L}^{j})$ 
where $n_{ij} = 2$ for 2-mode combinations involving two different modes, and 1 for the 
first harmonic of a mode; the relative phase of combination modes is defined as 
$\Delta\Phi_C = \Phi_L^{i\pm j} - (\Phi_L^{i} \pm \Phi_{L}^{j})$.}
\end{centering}
   \end{table*}

\subsection{Velocity curve and significance of detections}
\label{sec:velocities}

In order to search for any variations in the line-of-sight velocity we 
cross-correlated our 379 usable spectra using the mean 
spectrum as a template. Having thus constructed a velocity curve 
(Fig. \ref{fig:ligovel}b) we attempted to fit it in exactly the same way as 
described above, except that we fixed the frequencies to those derived from 
the light curve (Table \ref{gdtab}) leaving the amplitudes and phases free to
vary. A low-order polynomial was included in the fit in order to remove any 
slow variations. Given the presence of noise especially at the low frequency 
end, this imposition of frequencies from the light curve is a use of additional 
independent information, and merely reflects our (reasonable) assumption that we 
expect the same frequencies to be present in the velocity curve as those found in 
the light curve. 

As our spectra were taken through a wide slit (to preserve photometric
quality), the positions of the spectral lines depend on the exact position 
of GD 358 in the slit. Although this position should be tagged to the
position of the guide star, we found that additional random jitter
(e.g. due to guiding errors, windshake) was present.
An estimate of the scatter in the measured velocities due to wander can be 
obtained from shifts in the slit of the spatial profile under the assumption that 
these shifts are also representative of the scatter in the dispersion direction.
We fitted Gaussians to the spatial profiles and took the standard deviation
of the centroid positions as a proxy for the scatter; we find a value of
9\,\kms\ at a representative wavelength of 4471\,{\AA}. Including this value 
in the least squares fit of the velocity curve results in a $\chi_{\mathrm{red}}^{2}
\simeq 1$. The velocity amplitudes and phases we find are listed in Table 
\ref{gdtab}.

As a cross-check, we fitted 13 of the strongest lines (marked in Fig. \ref{fig:avspec})
in all 379 spectra with a combination of a Gaussian and a line or a 2$^{\mathrm{nd}}$-order 
polynomial to represent the continuum and constructed an average velocity curve using the 
Doppler shifts of all 13 lines. Typical uncertainties in the fitting of the line profiles 
are of the order of 14\,\kms\ at $\lambda$4471\,{\AA}. We then fit this velocity curve in 
exactly the same manner as described above and found good agreement (within the quoted
errors) with both the velocity amplitudes and phases derived from the cross-correlation 
procedure.

For the DAVs, the derivation of the velocity curve using the cross-correlation
technique is not reliable \citep[see the discussion in][]{vkcw:00} due to changes 
in the continuum slope during the pulsations -- an effect that is not taken into 
account in the cross-correlation procedure. The higher effective temperatures of 
the DBVs however, means that the continuum slope is not severely affected and 
therefore does not bias the Doppler shifts, making the procedure more reliable.

\begin{figure}[!t]
\plotone{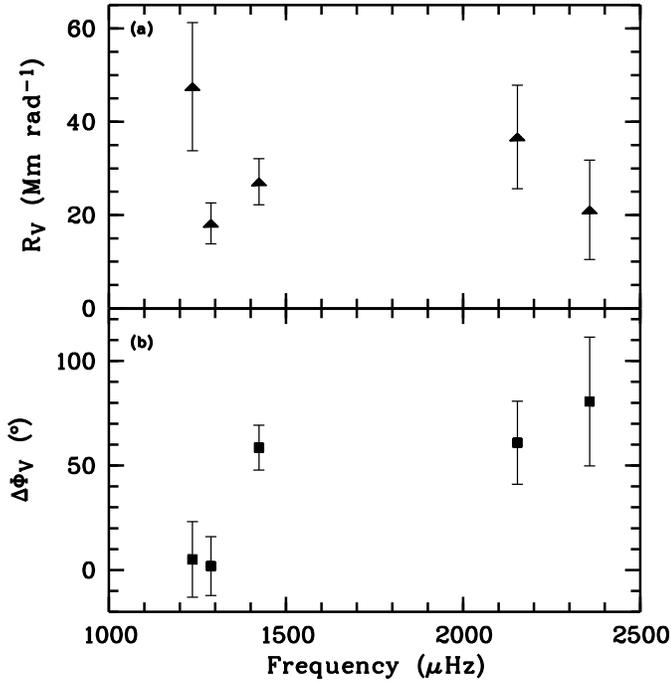}
\caption{\textbf{(a)} The relative velocity to light amplitude $R_V = A_V/2\pi fA_L$
and \textbf{(b)} $\Delta\Phi_V$, the phase difference between velocity and light, for
the 5 real modes.}
\label{fig:rvphiv}
\end{figure}

In order to obtain quantitative estimates of the significance of our measured
velocity amplitudes and to ascertain the contribution of random noise peaks, we 
conducted a simple Monte Carlo test: we randomly shuffled the velocities with
respect to the observation times and fit the resulting velocity curve in 
exactly the same way as the observations (as described above). We repeated
the procedure 1000 times and counted the number of times a peak larger
than that observed was found exactly at the frequency of each of the real modes.
We found that the peak at F2 had a $\simlt0.1$\% chance of being a random peak
while those at F1, F3, and F4 had a likelihood of 2\%, 5\%, and 9\% respectively 
of being chance occurrences. All other peaks were found to be insignificant.
We conclude therefore, that the modulations we see in the Doppler shifts are
due to intrinsic processes and that these are line-of-sight velocity variations 
associated with the pulsations. In what follows, we include all real modes and
treat the velocity amplitudes of those that were only marginally significant as 
upper limits.

We note as an aside that \citet{prov:00} invoke the velocites associated with the 
oscillatory motions to explain the width ($v\sin i \sim 60$\,\kms) of the 
\ion{C}{ii}\,$\lambda$\,1335\,{\AA} line. Our measured velocities do not support 
this speculation. The interpretation of the broad and shallow profiles as being 
due to a high rotational velocity is also at odds with that inferred from the 
rotationally-induced splittings \citep[0.9 days for the core to 1.6 days for 
the envelope,][]{wing:94}. Interestingly, several slowly rotating DAVs too show 
broad and flat (H$\alpha$) line cores, although this observation is not unique to 
the pulstors \citep{dkrot:98}.

\begin{table}[!t]
\caption[]{Comparison with the WET data sets}
\fontsize{7.5}{9.2}\selectfont
 \begin{centering}
 \setlength{\extrarowheight}{1.pt}
 \begin{tabular}{lclcccccc}
 \label{tab:gdwet}
       &             &      &       &         &        &        \\
\hline
       & Period      &  $n$ &  $m$  & WET 90  & WET 94 & Keck 99 \\
\multicolumn{4}{c}{} &
\multicolumn{3}{c}{Amplitude} \\
       & (s)         &      &       &  (\%)     &  (\%)    & (\%)  \\
\hline
 F1    &  776.42   &  17     & $-$1,(0)  & 0.49 (1.45$^*$) & 0.64 (2.21$^*$) &  3.10  \\
 F2    &  702.39   &  15     & \phs0     & 1.9$^*$         & 1.66            &  2.29  \\
 F3    &  809.28   &  18     & \phs0     & $\sim0.3^*$     & 1.36$^*$        &  1.07  \\
 F4    &  464.44   & \phn9   & \phs0,1   & 0.45$^*$,0.27   & 0.48$^*$,0.27   &  0.66  \\
 F5    &  424.26   & \phn8   & $-$1,0    & 0.49,0.5$^*$    & 0.45,0.92$^*$   &  0.62  \\ 
\hline
\end{tabular}
\tablecomments{The WET 90 and WET 94 results are from \citet{wing:94} and \citet{vuille:00}. 
respectively; all modes listed here were identifed in these studies as having $\ell=1$. 
Note that small (0.03 - 2\,$\mu$Hz) frequency shifts -- of unknown origin -- were found 
between the two WET data sets. The $n$ and $m$ values listed here are simply chosen from 
\citet{vuille:00} that are closest to our measured values. As the frequency splitting
decreases with decreasing $n$, the difficulty in discriminating between the various 
multiplet components within a triplet observed in the WET data and our single mode, 
increases. An asterisk next to the mode amplitude indicates that this mode had the 
largest amplitude in the triplet. We remind the reader that our amplitudes are a blend 
of the different $m$ components. For the $n=17$ triplet, we indicate the $m$ component 
that had the largest amplitude in both WET datasets in brackets for comparison purposes. 
Note that for some reason, \citet{wing:94} do not list the amplitude for the $n=18$ mode 
in their Table 2 even though it is clearly present and shows many combinations. The value 
above has been obtained by reading off the value from their Fig. 3.} 
\end{centering}
   \end{table}

\section{Real modes}

Following \citet{vkcw:00}, we adopt $R_V = A_V/2\pi fA_L$ and $\Delta\Phi_V 
= \Phi_V-\Phi_L$ as measures of the relative velocity to flux amplitude ratio 
and phase difference, respectively, for the real modes. We list these values in 
Table \ref{gdtab}, and plot them as a function of mode frequency in Fig. 
\ref{fig:rvphiv}.

As a (real) mode propagates upward, its environment changes from being largely
adiabatic (velocity maximum lags flux maximum by 90\arcdeg), to one where 
non-adiabatic effects are important. If $R_V$ and $\Delta\Phi_V$ follow the 
expected trends, they can be used to constrain the properties of the outer 
layers of the white dwarf. We describe these trends below.

In the convection zone, flux attenuation increases with increasing mode frequency i.e., 
$A_L$ decreases. However, turbulent viscosity in the convection zone ensures negligible 
vertical velocity gradients, with the result that the horizontal velocities are effectively 
independent of depth within the convection zone \citep{brick:90,gw:99b}. Thus, the
ratio between the velocity and flux amplitudes ($R_V$) is expected to increase with 
increasing mode frequency. Fig. \ref{fig:rvphiv}a shows no such trend.

For the adiabatic case, the phase lag between velocity and light is expected to be 
90$\arcdeg$; non-adiabatic effects reduce this lag. $\Delta\Phi_V$ is therefore expected 
to lie between 0$\arcdeg$ and 90$\arcdeg$. This is indeed found to be the case 
(Table \ref{gdtab}). 
As higher frequency modes are increasingly delayed by the convection zone, 
$\Delta\Phi_V$ should tend to zero with increasing mode frequency. 
Although the error bars are large, Fig. \ref{fig:rvphiv}b shows a 
trend opposite to the one just described. It is worth noting that \citet{vkcw:00} 
also found some indication of such a trend in their analysis of ZZ Psc (a DAV type
pulsator).

\begin{figure*}[!t]
\begin{center}
 \mbox{
   \includegraphics[width=0.5\textwidth,clip=]{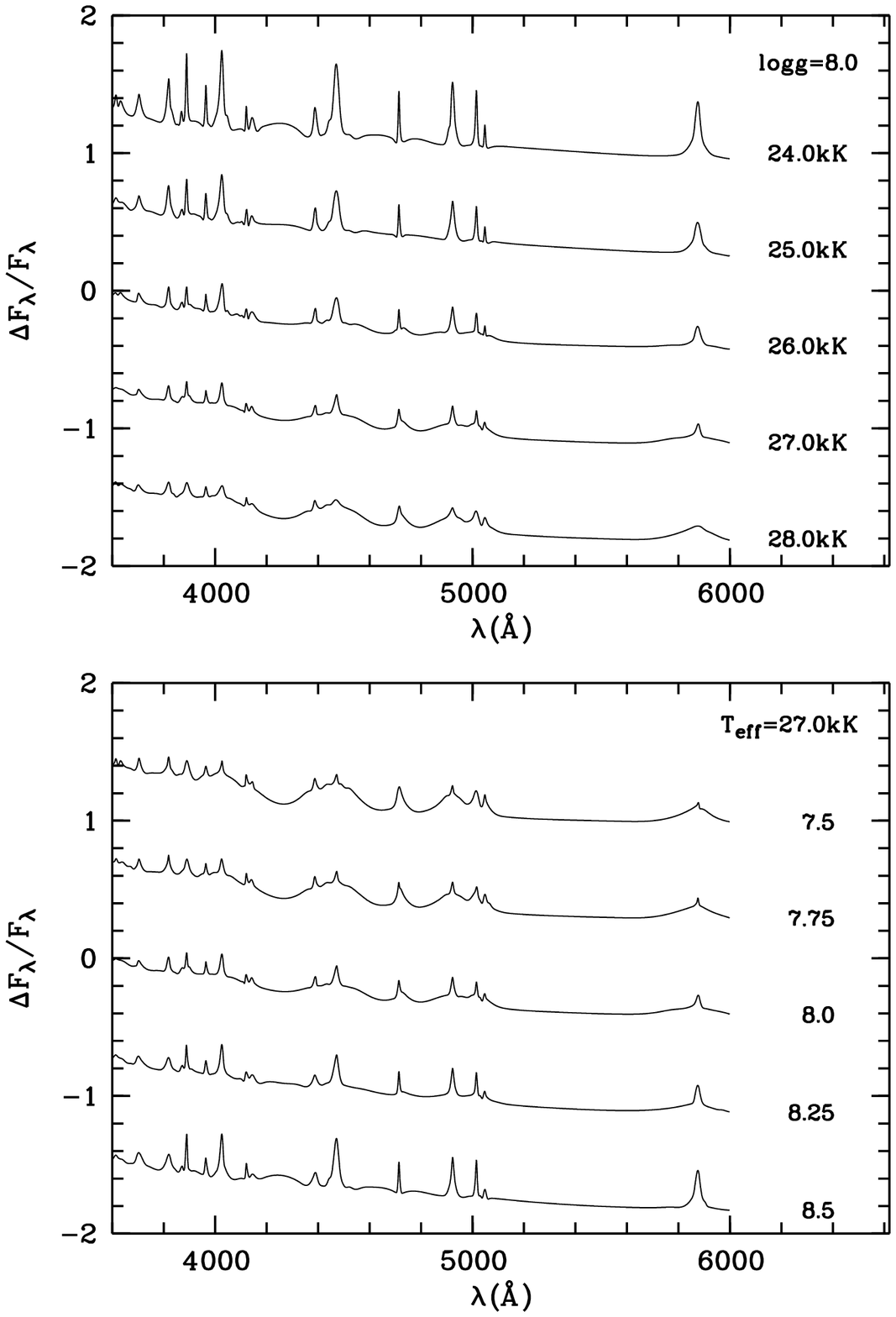}
   \includegraphics[width=0.5\textwidth,clip=]{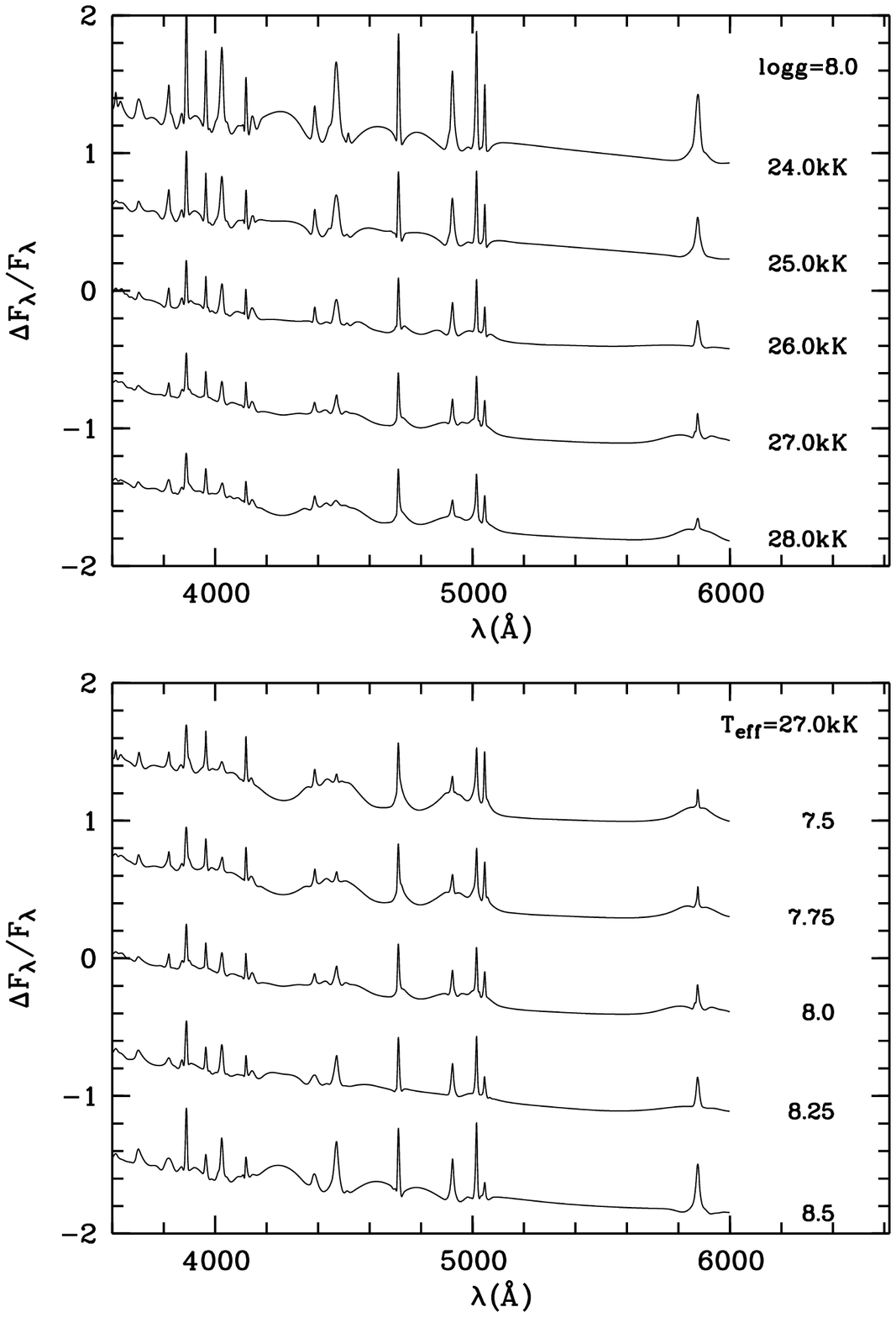}
      }
  \end{center}
\caption{\textbf{Top panel:} Model chromatic amplitudes (using ML2/$\alpha=
0.6$) for $\ell=1$ showing the effect of varying the effective temperature at
constant $\log g$ within the wavelength range of our observations.
\textbf{Bottom panel:} Same as above, but for varying $\log g$ at constant
effective temperature. The offset between the models is 0.7. All models have
been convolved with a Gaussian having a FWHM of 4.1\,{\AA} to emulate a
seeing profile. The panels on the left are for $\ell=1$ while those on the
right are for $\ell=2$.}
\label{fig:modelchamps}
\end{figure*}

Apart from the effect of increasing delay, another effect, that of increasing
non-adiabaticity with mode period \citep{gw:99b}, might also be at play. Indeed,
Fig. \ref{fig:rvphiv}b shows that the relative phase between velocity and flux
decreases as a function of mode period. It is difficult to estimate the relative 
importance the two effects.
Perhaps also of relevance to the above is the discontinous change in horizontal 
velocity across the boundary between the radiative interior and the convective 
layers \citep{gw:99b}. The shear layer at this boundary might be expected to impart
not insignificant phase shifts to the horizontal velocities since the spatial
acceleration due to viscous forces is roughly an order of magnitude larger
than the pressure perturbation for DAVs \citep[see Fig. 11 in][]{gautschy:96}. 
Whether this is indeed the case would have to be checked by investigating the behaviour 
of the modes in the presence of such a shear layer. It is not immediately evident 
from the description of \citet{gw:99b} how the phases of the horizontal velocities 
are affected across this jump as a function of mode period.

\begin{figure}[!t]
\plotone{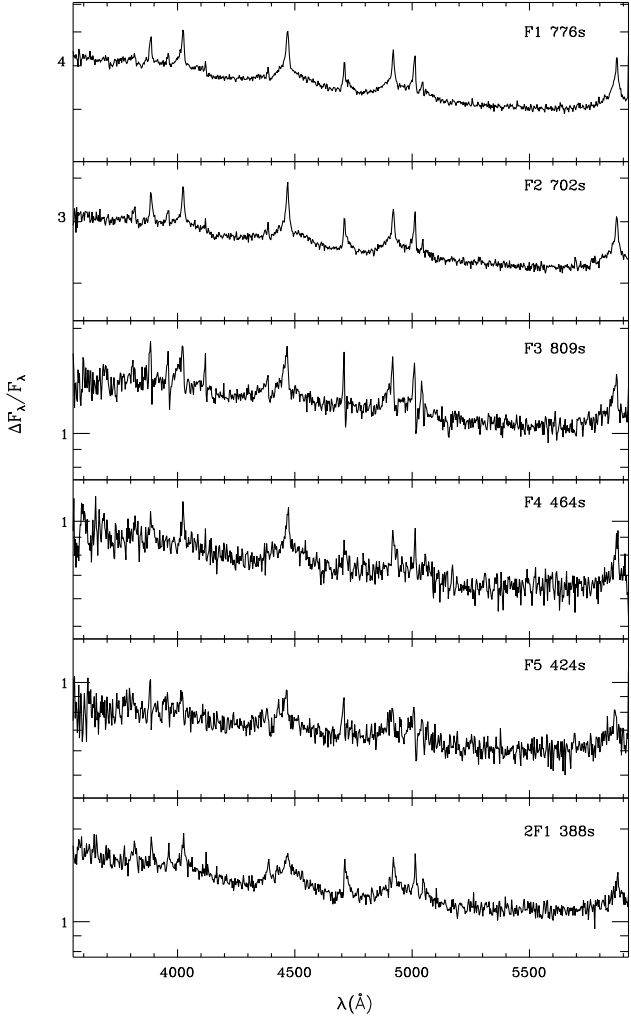}
\caption{Observed chromatic amplitudes, calculated in 3\,{\AA} wide bins,
shown here for the 5 real modes and the strongest combination mode. The periods
are indicated next to the names. }
\label{fig:gdchamp}
\end{figure}

\begin{figure}[!t]
 \plotone{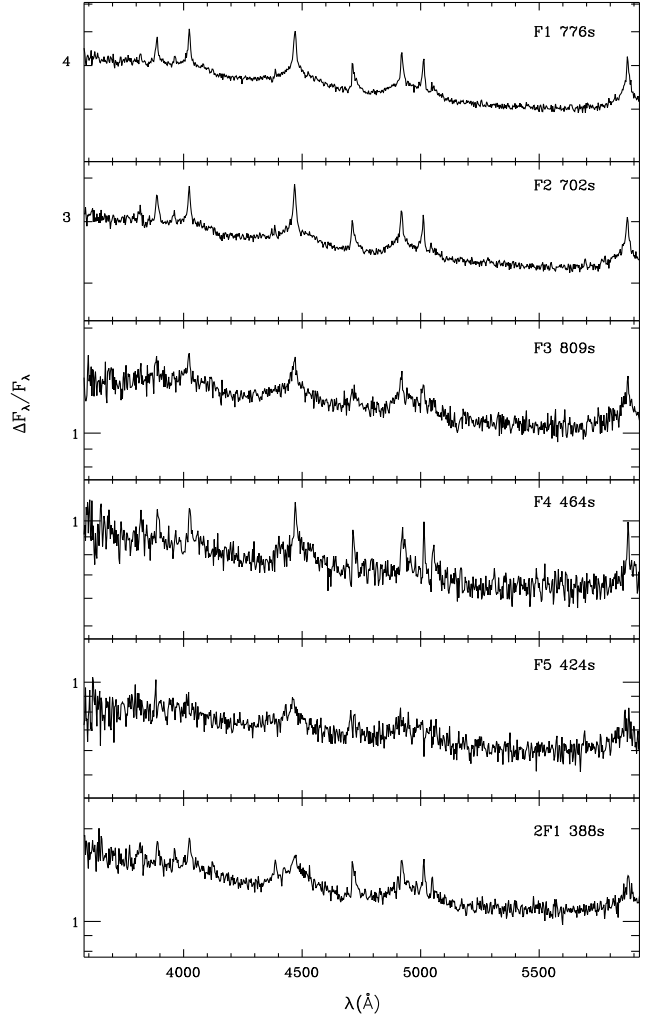}
 \caption{Same as Fig. \ref{fig:gdchamp}, but showing the effect of removing the 
  average Doppler shift on the chromatic amplitudes for F1 -- F5.
  The effect is most clearly seen e.g. in the $\lambda$4713\,{\AA} line of F3.
  Note the lack of significant change (c.f. Fig. \ref{fig:gdchamp}) for 2F1.}
 \label{fig:shiftch}
\end{figure}

\section{Chromatic amplitudes ($\mathbf{\Delta F_{\lambda}/F_{\lambda}}$)}
\label{sec:champ}

As mentioned in Sect. \ref{sec:intro}, the fractional pulsation amplitudes 
calculated as a function of wavelength bear an $\ell$-dependent signature. 
Even though all non-radial modes suffer from cancellation which occurs as 
a result of observing disc-integrated light, the cancellation increases with
increasing $\ell$. As limb-darkening increases at shorter wavelengths, the
effect of this cancellation is reduced, resulting in a net increase of 
pulsation amplitudes (for $\ell \le 3$). A similar effect is also apparent 
at optical wavelengths, especially in the absorption lines.
Comparing fractional amplitudes has the added advantage that uncertainties 
in the data acquisition and reduction processes are expected to cancel out; 
any differences between the synthetic and observed chromatic amplitudes are
therefore intrinsic.

The model atmospheres described in Sect. \ref{sec:modatm} have intensities
tabulated at nine different limb angles. We used these to compute a grid
of synthetic chromatic amplitudes by integrating over these intensities, 
weighted by Legendre polynomials which describe the surface distribution of 
the temperature perturbations. The change in intensity with respect to 
temperature (as a function of the limb angle) was computed using adjacent 
models of different effective temperature.

The effect of varying the effective temperature and $\log g$ is shown 
in Fig. \ref{fig:modelchamps}. Contrary to the case of the DA pulsators 
\citep{cvkw:00}, the models display a remarkable sensitivity 
to small changes in these model parameters. If the models are able to
match the observations, this sensitivity could be exploited to provide a
new way of measuring \Teff\ and $\log g$. The differences between $\ell=1$ 
and $\ell=2$ in the models are similar to the DA case, in that the continuum 
between the lines is more curved and the line cores sharper for the latter 
case.

\begin{figure}[!t]
\plotone{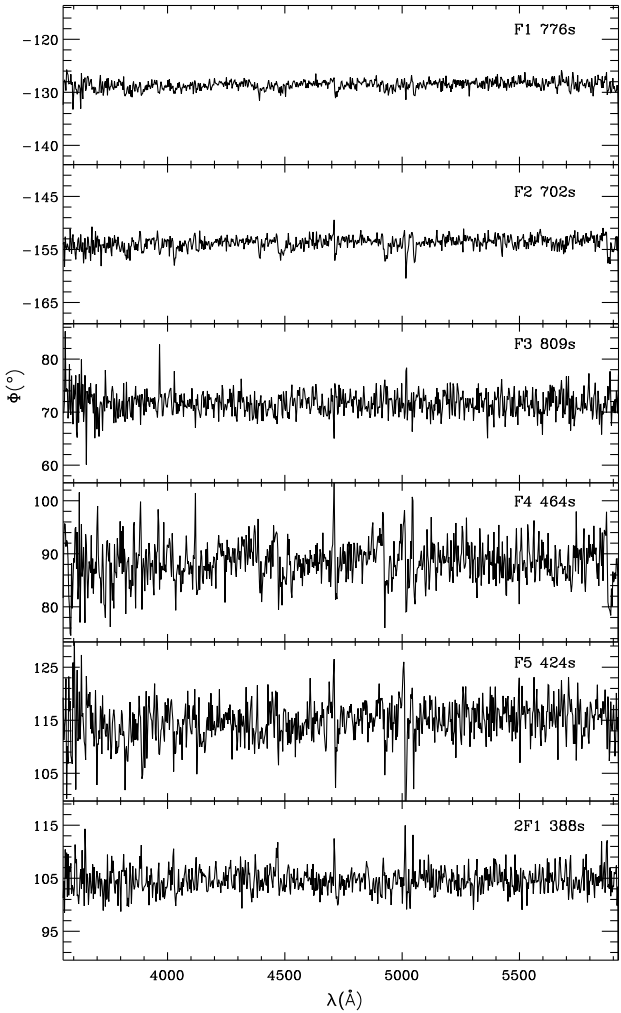}
\caption{Observed chromatic phases shown here for all real modes and the 
    strongest combination mode.}
\label{fig:gdph}
\end{figure}

To compare with models, we constructed a 2-D (wavelength-time) stacked image 
from our spectra and calculated the fractional pulsation amplitudes as a function 
of wavelength by fitting for the amplitudes and phases of the modes listed in 
Table \ref{gdtab}, the frequencies being held fixed to the tabulated values. 
The resulting chromatic amplitudes and phases for all real modes are shown 
in Figs. \ref{fig:gdchamp} and \ref{fig:gdph}.

\citet{cvkw:00} were able to assign $\ell$ values to the real modes in ZZ Psc
by comparing the observed chromatic amplitudes with each other. As they showed,
simple inspection of the chromatic amplitudes, especially when modes with different 
$\ell$ values are present, works well. This is particularly true when the target
is bright. Before comparing the chromatic amplitudes of GD 358 with the models, we 
note that all the real modes have the same general shape to one another implying that 
they share the same $\ell$ value (especially F1 and F2 which are nearly identical); 
given the WET results, this is most likely $\ell=1$. 

The only exception to the above is F3 (Fig. \ref{fig:gdchamp}) which exhibits
some differences compared to the other modes, most notably in the shape and
strength of the line cores at $\lambda$4120, 4713, and 5047\,{\AA}. The first and
last of these three lines are barely discernible in the chromatic amplitudes of the
other four real modes. Although the amplitudes of the line cores are in general greater
for $\ell=2$ than $\ell=1$ modes, it is difficult to disentangle the effect of a different
spherical degree from the dependence on the effective temperature and $\log g$. 
We also note that F3 has the highest velocity to flux amplitude ratio ($R_V$). 
Even though the differences are admittedly not significant compared to the other modes,
a higher $R_V$ can be indicative of a higher $\ell$ value as the light variations for such
modes suffer greater cancellation than the horizontal velocity variations \citep{dz:77}.

While the difference in appearance is significant, one has to be
careful in assigning a cause. In principle, line-of-sight variations
that are in phase with the light curve, could induce differences in
relative amplitude as well \citep[see][]{cvkw:00}. These scale
with $dF_\lambda/d\lambda$ and thus should be most pronounced near
sharp lines. This is indeed seen for F3, which additionally has the
highest value of $R_V\cos\Delta\Phi_V$, i.e., for F3 the line-of-sight
velocity variations are expected to have the largest effect on the
chromatic amplitudes. In order to test whether this was indeed the
underlying cause of the difference in appearance, we removed the
Doppler shift from all the spectra (using the average velocities shown 
in Fig. \ref{fig:ligovel}) and recomputed the chromatic amplitudes.  
The result is shown in Fig. \ref{fig:shiftch}. Clearly, having taken out 
the effect induced by the line-of-sight velocity variations, F3 becomes 
very similar in appearance to F1 and F2. On the basis of this, we conclude 
that F3 too shows $\ell=1$-like behaviour. This is consistent the WET 1994 
data which shows a triplet at this period, indicative of an $\ell=1$ 
mode.\footnote{The WET 1990 data only show one low amplitude peak at 809\,s.}
We stress, though, that the above leaves open the question of why F3 has
a larger $R_V\cos\Delta\Phi_V$.

It is interesting to note that the chromatic amplitude of the strongest
combination mode, 2F1, is different in appearance compared to the five
real modes, in that it has a larger curvature in the continuum between
the lines cores (Fig. \ref{fig:gdchamp}). This is probably due to a dominant 
$\ell=2$ component (see Fig. \ref{fig:modelchamps}).

Incidentally, this is to be expected if our match -- based on \citet{vuille:00}
-- of $m=-1$ is correct (Table \ref{tab:gdwet}), since a combination of 
$Y_{1}^{-1}Y_{1}^{-1}$ (where the $Y_{\ell}^{m}$ are spherical harmonics)
has a distribution described by $Y_{2}^{-2}$ only. 
We cannot, however, definitively rule out the possibility that F1 has $m=0$
or $m=1$ which would mean the presence of a $Y_{0}^{0}$ component in the
former case in addition to the $Y_{2}^{0}$ component.
Unfortunately, the weakness of the other combinations means that we 
cannot subject them to a meaningful test.

\begin{figure*}[!ht]
 \begin{center}
  \mbox{
   \includegraphics[width=0.5\textwidth,clip=]{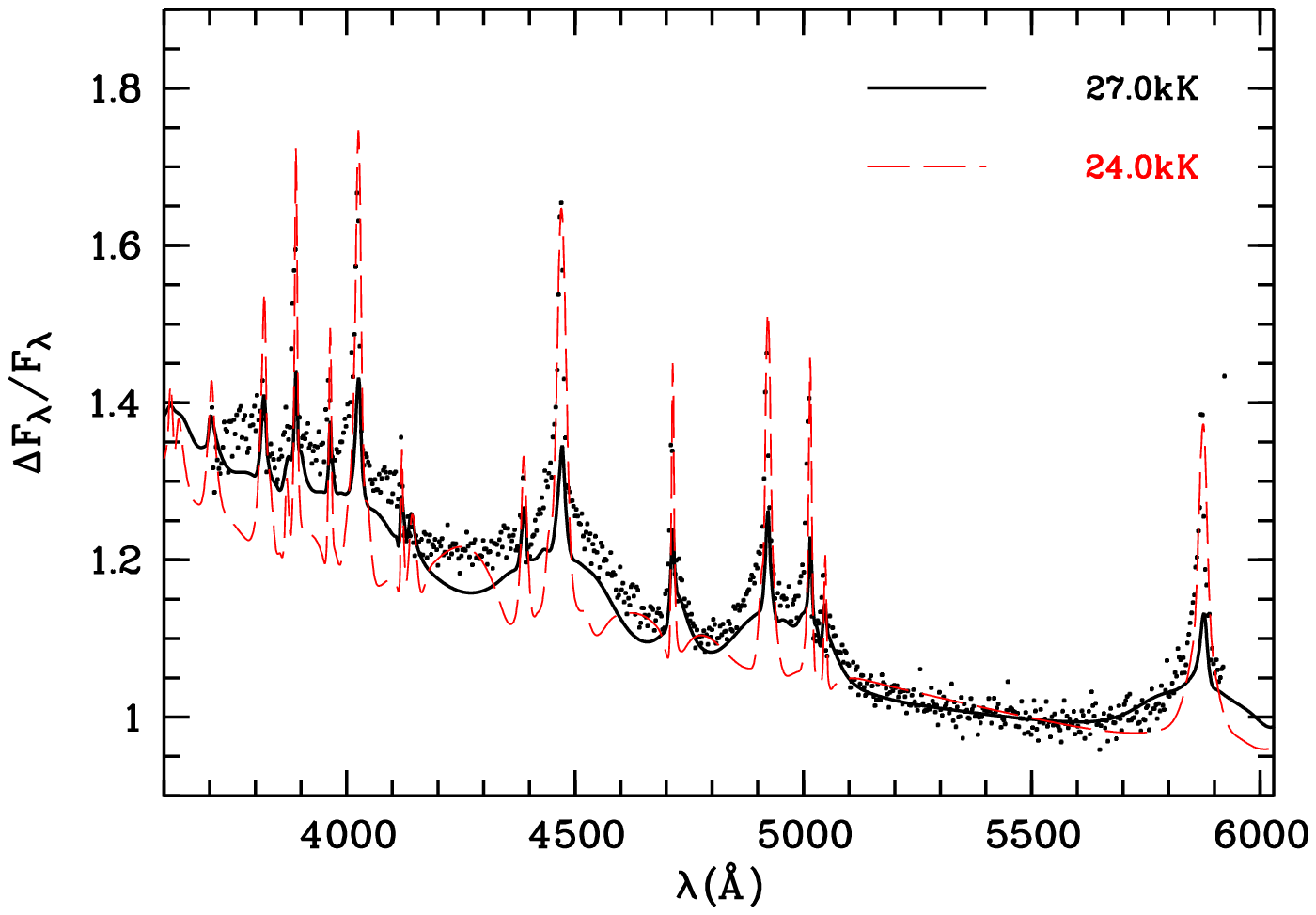}
   \includegraphics[width=0.5\textwidth,clip=]{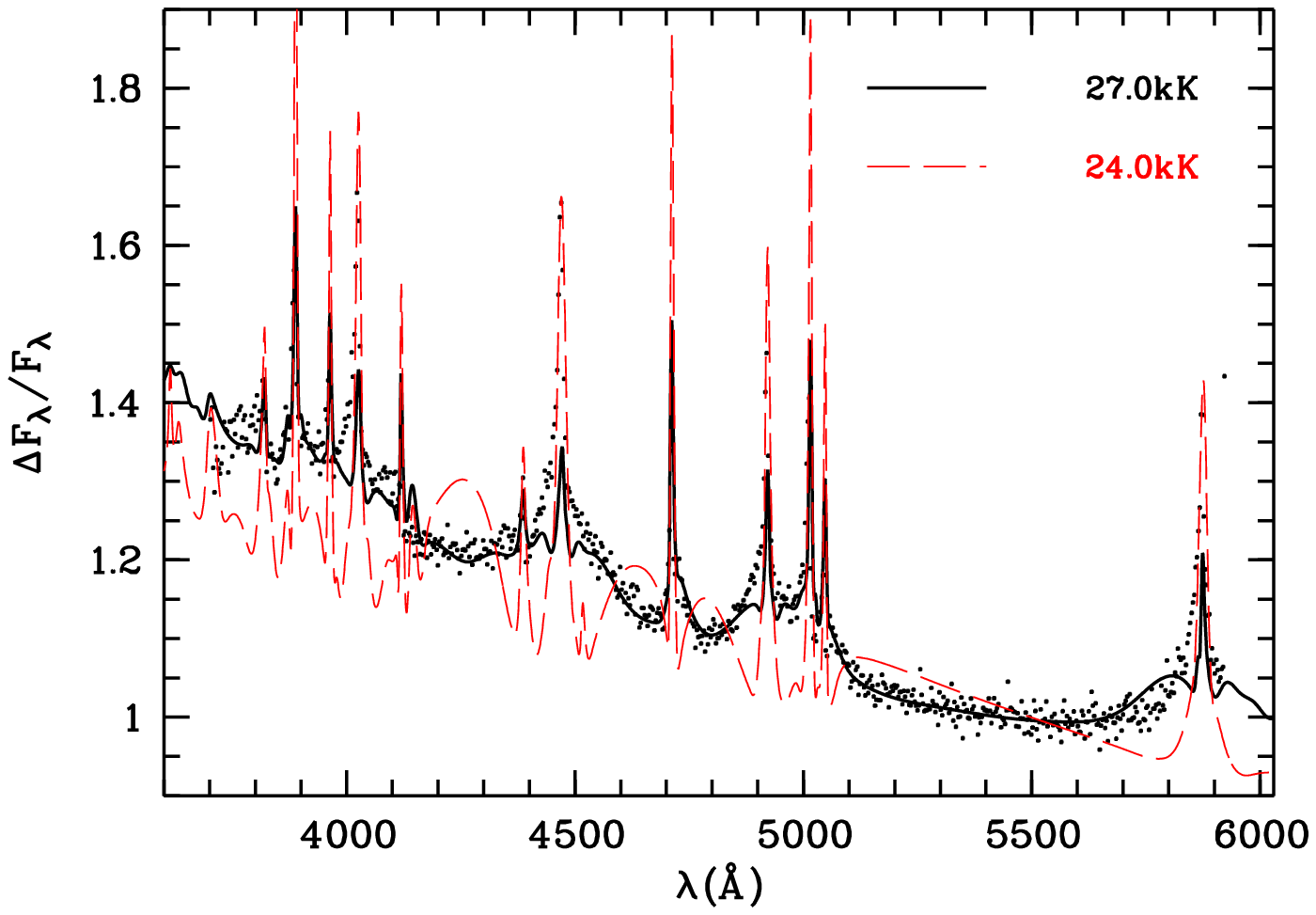}
       }
   \end{center}
   \caption{\textbf{Left:} The observed chromatic amplitude for the mode having the
           largest amplitude flux variations, F1, (dots) overlaid with models calculated
           at the two different effective temperatures with $\log g=8$ for $\ell=1$ modes
           and with intermediate convective efficiency.
           \textbf{Right:} For models having $\ell=2$ }
  \label{fig:modcfdata_ml2_06}
 \end{figure*}

 \begin{figure*}
 \begin{center}
   \mbox{
   \includegraphics[width=0.5\textwidth,clip=]{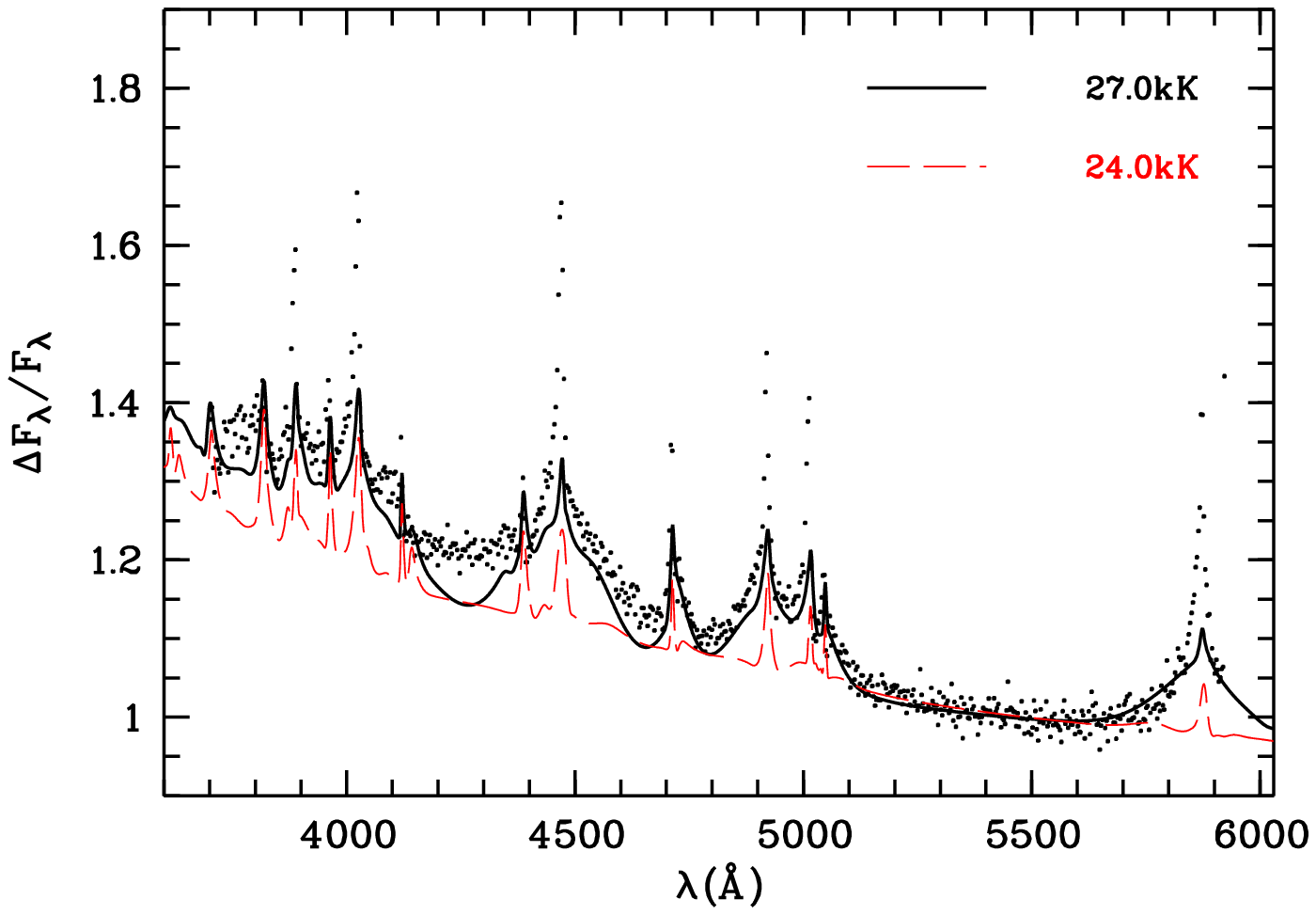}
   \includegraphics[width=0.5\textwidth,clip=]{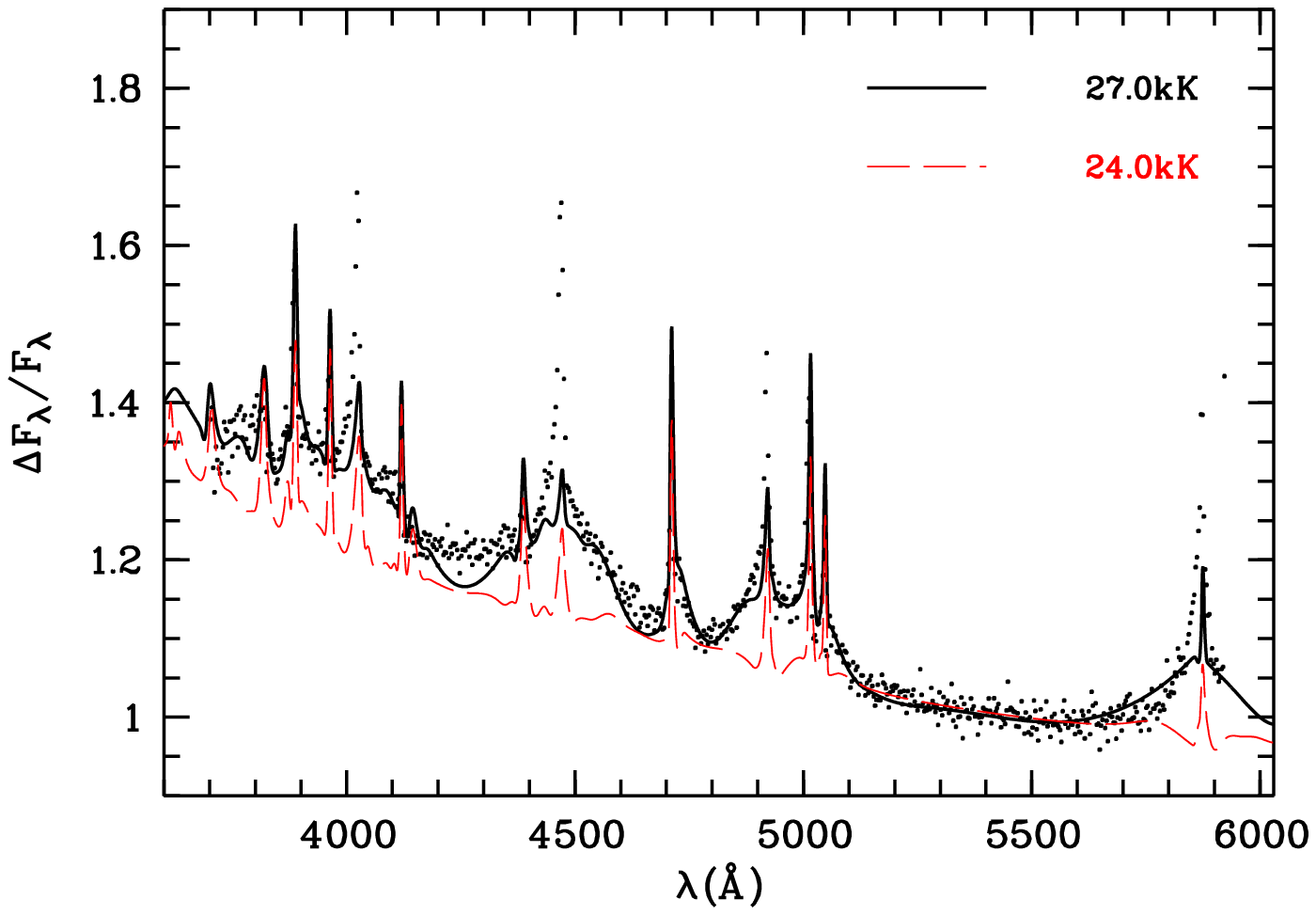}
        }
  \end{center}
\caption{\textbf{Left:} Same as for Fig. \ref{fig:modcfdata_ml2_06}, but with
         the models calculated using a lower convective efficiency (ML1/$\alpha=0.5$). 
         For $\ell=1$. \textbf{Right:} For $\ell=2$.}
\label{fig:modcfdata_ml1_05}
\end{figure*}

Although our fit to the mean spectrum (Fig. \ref{fig:stackmodavspec}) is arguably 
reasonable, we find that for the effective temperature derived from the spectrum
(24\,kK), the model chromatic amplitudes do not even look qualitatively similar to
the observations. Fig. \ref{fig:modcfdata_ml2_06} shows the chromatic amplitude of 
the strongest mode (F1) compared to synthetic chromatic amplitudes for two effective 
temperatures each for $\ell=1$ and $\ell=2$. We can only match the shape of the 
pseudo-continuum between the line cores if we use models with higher ($\sim$\,27\,kK) 
temperatures.

We do not expect to fit the line cores as these are probably formed
in parts of the atmosphere where $\tau\ll1$ and where, as a result,
deviations from LTE might be important. The (pseudo-)continuum,
however, should not be affected and we stress that it is these regions 
of the chromatic amplitudes that we seek to match.

We see that the observed wings of the lines are broader than those
of the models calculated using an intermediate convective efficiency
(ML2/$\alpha=0.6$). Less efficient convection would make the temperature
gradient steeper. To test whether the temperature stratification of the models 
is the culprit, we calculated models in which the convective efficiency was three 
times lower. In Figs. \ref{fig:modcfdata_ml2_06} and \ref{fig:modcfdata_ml1_05} 
we compare models having different convective efficiencies to the data and see 
that there is a slight improvement to the match in the wings of the lines for 
the model having a lower convective efficiency (ML1/$\alpha=0.5$) but that the 
match in some of the continuum regions has worsened. It may be possible to reproduce 
the chromatic amplitudes with a different choice of mixing length or mixing-length 
prescription. However, we feel this is unlikely, as we failed to reproduce the 
chromatic amplitudes for ZZ~Psc even with a very extensive set of models.  
We will return to this briefly below.

Phase changes within the lines due to the line-of-sight velocity variations
are apparent (Fig. \ref{fig:gdph}) especially for the stronger lines.
As in previous studies of this nature of DA pulsators, we find a very slight 
($\sim$1\arcdeg) slope in phases over the wavelength range although this is much
less obvious than was found for at least two DAVs \citep{vkcw:00,kotakhs:02}.

\section{Discussion and conclusions}
\label{sec:conc}

As outlined at the outset, our main aims were to investigate the 
pulsation properties of GD 358, to compare these with the better studied
DAVs, and to use our data to place constraints on model atmospheres.

We began by determining the modulations present in the light curve and found
that all five real modes agreed very well with previous WET observations.

We also found variations in the Doppler shifts of the spectral lines that 
were coincident with those determined from the light curve. We concluded that 
these were line-of-sight velocity variations associated with the largely 
horizontal motion of the pulsations. 

Using the velocity to light amplitude ratios ($R_V$) and phases ($\Delta\Phi_V$) 
we tested theoretically expected trends. We found no evidence for a general increase 
in $R_V$ with mode frequency. This has also been the case for the DAVs.
Although $\Delta\Phi_V$ was found to lie between 0 and 90\arcdeg, as expected
we found some evidence that it increases with mode frequency -- a trend opposite 
to that predicted. Previous measurements of the DAVs \citep{vkcw:00} have also 
found marginal evidence for such a trend. 

The wavelength-dependent fractional pulsation amplitudes proved to be invaluable 
in two altogether different ways. First, the resemblance of most of the observed
chromatic amplitudes to each other led to the conclusion that they all shared the 
same spherical degree, most likely, $\ell=1$.  
Our results provide an entirely independent confirmation of the WET results.

We also showed that the chromatic amplitude of 2F1 was different, and bore the 
signature of an $\ell=2$ mode. Our $m$ identification -- based on the WET 1994 results
-- suggests that F1 has $m=-1$. If true, it implies a redistribution of power between 
the $m$ components in that triplet, which previously showed a dominant $m=0$ component
(see Table \ref{tab:gdwet}).

Secondly, although our synthetic chromatic amplitudes failed to reproduce the 
observations for any combination of \Teff\ and $\log g$, they unambiguously 
indicated that a temperature higher than that inferred from our mean spectrum 
(\Teff =24\,kK; $\log g$=8) was a better match at all wavelengths.
Thus while our average spectrum favours a lower temperature, the chromatic
amplitudes indicate a higher temperature, in agreement with temperature 
determinations based on ultra-violet spectra. The cause of this discrepancy
remains unexplained. 

We also found that the chromatic amplitudes could be reproduced only
qualitatively, with the models failing to reproduce both the cores and
the wings of the lines. Slight improvement is obtained -- mainly in
the wings of the lines -- with models of lower convective efficiency
than ML2/$\alpha=0.6$. In detail, however, the models still do not
match, as is also the case for the DAV ZZ~Psc \citep{cvkw:00}
even when trying models for a much more extensive set of mixing-length
parameters. The above may mean mean that there is an intrinsic
problem with the atmospheric models, e.g., that the real temperature
structure simply cannot be reproduced accurately enough with the
mixing-length approximation. If so, perhaps the observations can be
used to derive the temperature stratification empirically.

Alternatively, there may be a problem with the way that the
atmospheric models are used to calculate model chromatic amplitudes.
In particular, our calculations assume that the temperature
variations associated with the pulsations are described well by a
single spherical harmonic, and we take into account only the first
derivative of the flux with respect to temperature. The latter effect
is rather small, but the former may be important: \citet{ik:01}
used numerical simulations to investigate these issues and
found that as the amplitude of a mode increases, the surface flux
distributions do deviate more and more from spherical harmonics. 
The effects become pronounced for pressure perturbations
larger than about 10\%, which correspond to visual flux variations
with amplitudes of about 5\% for DAV and 3\% for DBV, i.e., at about 
the level observed for the strongest modes.

No direct comparison of these models with observations has yet been made, 
but an empirical estimate can be made by considering all the non-linearities 
in terms of combination frequencies. Clearly, second-order corrections 
cannot influence the chromatic amplitudes of a real mode, since these
corrections appear as harmonics, sums, and differences of the real
modes. Third-order corrections, however, will have terms -- such as
2F1${}-{}$F1 and F1+F2$-$F2 -- which coincide with the real modes,
but are described by surface distributions with different $\ell$ and
can thus influence the shape of the chromatic amplitudes measured at
the frequency of the real mode. Indeed, \citet{vuille:00} find a large 
number of third-order combination frequencies for GD~358, with
amplitudes up to a quarter of those of the real modes. While this is
a significant fraction, one should bear in mind that their surface
flux distribution will be dominated by their $\ell=0$, 1, and 2
components, which will not change the inferred chromatic amplitudes
much. It thus remains unclear whether or not these higher-order terms
can be responsible for the mismatch between the observed and model
chromatic amplitudes.

Fortunately, there is a clear prediction: for low-amplitude pulsators,
which show no or only very weak combination modes, the chromatic
amplitudes should not be influenced by higher-order corrections. 
Thus, if the chromatic amplitudes of such pulsators can be matched
by models, higher-order corrections are at play for the mismatches 
observed so far. If they do not, the problem must be one intrinsic to
the model atmospheres. With accurate observations of low amplitude DBVs
and detailed models, we can expect rapid progress.

\begin{acknowledgements}
We thank the referee, G. Handler, for his positive comments.
R.K. would like to sincerely thank H-G. Ludwig for many discussions.
M.H.vK acknowledges support for a fellowship of the Royal Netherlands Academy of 
Arts and Sciences. This research has made use of the SIMBAD database, operated at 
CDS, Strasbourg, France. 
\end{acknowledgements}

\end{document}